\documentclass[hyper]{JHEP} 

\usepackage{epsfig}

\newcommand\fverb{\setbox\pippobox=\hbox\bgroup\verb}

\newcommand\fverbdo{\egroup\medskip\noindent%

            \fbox{\unhbox\pippobox}\ }

\newcommand\fverbit{\egroup\item[\fbox{\unhbox\pippobox}]}

\newbox\pippobox

\title{Remark About Hamiltonian Formulation of
 Non-Linear Massive Gravity in
St\"{u}ckelberg Formalism}
\author{J. Kluso\v{n}\\
Department of
Theoretical Physics and Astrophysics\\
Faculty of Science, Masaryk University\\
Kotl\'{a}\v{r}sk\'{a} 2, 611 37, Brno\\
Czech Republic\\
E-mail: \email{klu@physics.muni.cz}}
\preprint{}

 \abstract{We perform the Hamiltonian
 analysis of the specific model of the non-linear
 massive gravity in
 St\"{u}ckelberg formalism where the square root structure
 is replaced by introducing
 auxiliary fields. Following
 arXiv:1203.5283 [hep-th] we show that given
 theory possesses an additional primary constraint. Then we perform
analysis of the consistency of all constraints during the time
development of the system and we argue for the existence of the
additional constraint. As a result the theory possesses the
correct number of physical degrees of freedom corresponding
to the massive gravity.}
 \keywords{Massive Gravity}

\def\bA{\mathbf{A}}

\def\mC{\mathcal{C}}
\def\be{\begin{equation}}

\def\ee{\end{equation}}

\def\bea{\begin{eqnarray}}

\def\eea{\end{eqnarray}}

\def\tr{\mathrm{tr}\, }
\def\tmH{\tilde{\mH}}

\def\mH{\mathcal{H}}

\def\tr{\mathrm{Tr}}

\def\bx{\mathbf{x}}
\def\by{\mathbf{y}}

\newcommand{\hg}{\hat{g}}

\newcommand{\mK}{\mathcal{K}}
\newcommand{\hmK}{\hat{\mK}}

\newcommand{\mU}{\mathcal{U}}

\newcommand{\mG}{\mathcal{G}}

\def \bA{\mathbf{A}}

\newcommand{\bT}{\mathbf{T}}

\newcommand{\mL}{\mathcal{L}}

\def\pb #1{\left\{#1\right\}}

\begin{document}
\section{Introduction and Summary}\label{first}
One of the most challenging problem is
to find consistent formulation of
massive gravity. The first attempt for
construction of  this theory is dated
to the year 1939 when Fierz and Pauli
formulated its version of linear
massive gravity \cite{Fierz:1939ix}
\footnote{For review, see
\cite{Hinterbichler:2011tt}.}. However
it is very non-trivial task to find  a
consistent non-linear generalization of
given theory and it remains as an
intriguing theoretical problem. It is
also important to stress that recent
discovery of dark energy and associated
cosmological constant problem has
prompted investigations in the long
distance modifications of general
relativity, for review, see
\cite{Clifton:2011jh}.

It is natural to  ask the question
whether it is possible to construct
theory of massive gravity where one of
the constraint equation and associated
secondary constraint eliminates the
propagating scalar mode. It is
remarkable that linear Fierz-Pauli
theory does not suffer from the
presence of such a ghost. On the other
hand it was shown by Boulware and Deser
\cite{Boulware:1973my} that ghosts
generically reappear at  the non-linear
level. On the other hand it was shown
recently by de Rham and Gobadadze in
\cite{deRham:2010ik} that it is
possible to find such a formulation of
the massive gravity which is ghost free
in the decoupling limit. Then it was
shown in \cite{deRham:2010kj} that this
action that was written in the
perturbative form can be resumed into
fully non-linear actions \footnote{For
related works, see
\cite{Mirbabayi:2011xg,Sjors:2011iv,Burrage:2011cr,Gumrukcuoglu:2011zh,Berezhiani:2011mt,
Comelli:2011zm,Kluson:2011aq,Comelli:2011wq,Mohseni:2011vv,Gumrukcuoglu:2011ew,Hassan:2011zd,
Hassan:2011tf,D'Amico:2011jj,deRham:2011qq,deRham:2011pt,Gruzinov:2011mm,Koyama:2011yg,
Nieuwenhuizen:2011sq,
Koyama:2011xz,Chamseddine:2011bu,Volkov:2011an}.}.
The general analysis of the constraints
of given theory has been performed
 in \cite{Hassan:2011hr}. It was argued
 there that it is possible to perform
 such a redefinition of the shift
 function so that the
resulting theory still contains the
Hamiltonian constraint. Then it was
argued that the presence of this
constraint allows to eliminate the
scalar mode and hence the resulting
theory is the ghost free massive
gravity. However this analysis was
questioned in \cite{Kluson:2011qe}
where it was argued that it is possible
that this constraint is the second
class constraint so that the phase
space of given theory would be odd
dimensional. On the other hand in the
recent paper \cite{Hassan:2011ea} very
nice analysis of the Hamiltonian
formulation of the most general gauge
fixed non-linear massive gravity
actions was performed  with an
important conclusion that the
Hamiltonian constraints has zero
Poisson brackets. Then the requirement
of the preservation of this constraint
during the time evolution of the system
implies an additional constraint. As a
result given theory has the right
number of constraints for the
construction of non-linear massive
gravity without additional scalar mode
\footnote{For related work, see
\cite{Golovnev:2011nz}.}.

All these results suggest that the
gauge fixed form of the non-linear
massive gravity actions could be ghost
free theory. On the other hand the
manifest diffeomorphism invariance is
lacking and one would like to confirm
the same result in the gauge invariant
formulation of the massive gravity
action using the St\"{u}ckelberg
fields. In fact, it was argued in
\cite{deRham:2011rn} that for some
special cases  such a theory possesses
an additional primary constraint whose
presence implies such a constraint
structure of given theory that could
eliminate one additional scalar mode.
The generalization of given work to the
case general metric was performed
 \cite{Kluson:2011rt}. We
found the Hamiltonian form of given action and determined primary
and secondary constraints of given theory. Then it was shown
 that due to
the structure of the non-linear massive gravity action this theory
possesses one primary constraint that is a consequence of the fact
that $\det V^{AB}=\det (\partial_i \phi^A\partial_j \phi^B
g^{ij})=0$) as was firstly shown in \cite{Hassan:2012qv}
\footnote{We should stress that in the previous versions of given
paper the constraint $\det V^{AB}=0$ was not taken into account and
hence the  wrong conclusions were reached.}. It was also shown there that
this result has a crucial consequence for the structure of the
theory. On the one hand this constraint could provide a mechanism
for the elimination of an additional scalar mode however on the
other hand the condition $\det V^{AB}=0$ makes the calculation of
the algebra of the Hamiltonian constraints very difficult with
exception of two dimensional massive gravity.

These results suggest that it is still
very instructive to analyze the
non-linear massive gravity further in
order to fully understand to it. For
that reason we perform the Hamiltonian
analysis of the model of non-linear
massive gravity action in the form that
was presented in
\cite{Golovnev:2011nz,Buchbinder:2012wb}.
We consider this action written in
manifestly diffeomorphism invariant way
using the collection of the
St\"{u}ckelberg fields. We perform the
Hamiltonian analysis. As in
\cite{Hassan:2012qv} we determine the primary constraint
of the theory
\footnote{At this place we should again  stress that this
result is different from the conclusion that
was presented in the first version of this paper. In fact,
after publication of the paper
\cite{Hassan:2012qv} we recognized that we made
crucial mistake in the analysis of the properties of the matrix $V^{AB}=
g^{ij}\partial_i\phi^A\partial_j\phi^B$.}. We also show
that the Poisson brackets between the momentum conjugate to the auxiliary
fields which are the primary constraints and corresponding
secondary constraints possesses zero eigenvector. Then
we analyze the consistency of all these constraints with the
time evolution of the system and we argue for the existence
of the additional constraint. Finally we check again
consistency of all constraints with the time evolution of the system
and argue that given system of constraints is closed with following
picture. We have four first class constraints corresponding to
the spatial diffeomorphism and Hamiltonian constraints. We also have one
first class constraint that is the particular linear combination
of  the momentum conjugate to the auxiliary fields. By gauge fixing
we eliminate this momentum together with corresponding one auxiliary fields.
Then togehter with  18 the second class constraints we can completely eliminate
all auxiliary fields at least in principle. Finally we have two second
class constraints that can eliminate one scalar degree of freedom with corresponding
conjugate momenta. As a result the counting of the physical degrees of freedom
suggests that we have the correct number of physical degrees of freedom
of the non-linear massive gravity. In other words we showed the consistency of
this particular model of non-linear massive gravity.

The extension of this work is as follows when we would like to
extend given analysis to the most general form of the non-linear massive gravity.
This is much more difficult task due to the necessity to introduce additional
auxiliary fields. We hope to return to this problem in future.

\section{Non-Linear Massive Gravity}\label{second}
Let us consider following non-linear massive
gravity action \cite{deRham:2010kj}
\begin{equation}\label{Smassive}
S=M_p^2\int d^4x
\sqrt{-\hg}{}^{(4)}R(\hg)-
\frac{1}{4}M_p^2 m^2\int d^4x
\sqrt{-\hg}\mU (\hg^{-1}f) \ .
\end{equation}
Note that by definition $\hg^{\mu\nu}$
and $f_{\mu\nu}$ transform under
general diffeomorphism transformations
$x'^\mu=x'^\mu(x)$ as
\begin{equation}
\hg'^{\mu\nu}(x')= \hg^{\rho\sigma}(x)
\frac{\partial x'^\mu}{\partial x^\rho}
\frac{\partial x'^\nu}{\partial
x^\sigma} \ , \quad f'_{\mu\nu}(x')=
f_{\rho\sigma}(x)\frac{\partial
x^\rho}{\partial x'^\mu}\frac{\partial
x^\sigma}{\partial x'^\nu} \  .
\end{equation}
Now the requirement that the
combination $\hg^{-1} f$ has to be
diffeomorphism invariant implies that
the potential $\mU$ has to contain the
trace over space-time indices. Further,
it is convenient to parameterize the
tensor $f_{\mu\nu}$ using four scalar
fields $\phi^A$ and some fixed
auxiliary metric
$\bar{f}_{\mu\nu}(\phi)$ so that
\begin{equation}
f_{\mu\nu}=\partial_\mu\phi^A\partial_\nu
\phi^B \bar{f}_{AB}(\phi) \ ,
\end{equation}
where the metric $f_{AB}$ is invariant
under diffeomorphism transformation
$x'^\mu=x^\mu(x')$ which however
transforms as a tensor under
reparemetrizations of $\phi^A$. In what
follows we consider
$\bar{f}_{AB}=\eta_{AB}$, where
$\eta_{AB}=\mathrm{diag}(-1,1,1,1)$.

The fundamental ingredient of the
non-linear massive gravity is the
potential term. The most general forms
of this potential were derived in
\cite{Hassan:2011vm,deRham:2010kj}. Let
us consider the minimal form of the
potential introduced in
\cite{deRham:2010kj}
\begin{eqnarray}
\mathcal{U}(g,H)&=&
-4\left(\left<\mK\right>^2-\left<\mK^2\right>\right)
=\nonumber \\
&=&
 -4\left( \sum_{n\geq
1} d_n \left<H^n\right>\right)^2-
8\sum_{n\geq 2} d_n\left<H^n\right> \ ,
\nonumber \\
\end{eqnarray}
where we now have
\begin{eqnarray}
H_{\mu\nu}&=&\hg_{\mu\nu}-
\partial_\mu\phi^A\partial_\nu\phi^B\eta_{AB}
\ , \quad
H^\mu_\nu=\hg^{\mu\alpha}H_{\alpha\nu}
\ , \nonumber \\
\mK^\mu_\nu&=&\delta^\mu_\nu-\sqrt{\delta^\mu_\nu-
H^\mu_\nu}=-\sum_{n=1}^\infty d_n
(H^n)^\mu_\nu \ , \quad
d_n=\frac{(2n)!} {(1-2n)(n!)^24^n} \ .
\nonumber \\
\end{eqnarray}
and where $(H^n)^\mu_\nu=
H^\mu_{\alpha_1}H^{\alpha_1}_{\alpha_2}
\dots H^{\alpha_{n-1}}_\nu$.

As was observed in \cite{Kluson:2011rt}
 in order to find the Hamiltonian
formulation of given action it is more
convenient to consider the potential
term written as the trace over Lorentz
indices rather then the curved space
ones. Explicitly, we have
\begin{eqnarray}
\mU
=-4(<\hmK>^2-<\hmK^2>)= -4(\hmK^A_{ \
A})^2+4\hmK^A_{\ B}\hmK^B_{ \ A} \ ,
 \nonumber \\
\end{eqnarray}
where we defined
\begin{equation}
\hmK^A_{ \ B}=\delta^A_{ \
B}-\sqrt{\delta^A_{ \ B}-\bA^A_{ \ B}}
 \ .
\end{equation}
Even if it is possible to perform the
Hamiltonian analysis for general form of
the potential we now consider following
potential term
\begin{equation}\label{mass1}
\mL_{matt}=-2M_p^2 m^2\sqrt{g}N
\tr\sqrt{\bA} \ .
\end{equation}
Following
\cite{Golovnev:2011aa,Buchbinder:2012wb}
we introduce  auxiliary fields
$\Phi^A_{ \ B}$ and rewrite the
potential term into the form
\begin{equation}\label{mass2}
\mL_{matt}=
 -M_p^2m^2 \sqrt{g}N
(\Phi^A_{ \ B}+(\Phi^{-1})^A_{ \
B}\bA^B_{ \ A}) \ , \quad  \bA^B_{ \ A}=
\hg^{\mu\nu}\partial_\mu\phi^B\partial_\nu\phi_A
\ .
\end{equation}
Note that we have to presume an
existence of the inverse matrix
$(\Phi^{-1})^A_{ \ B}$. Then in order
to see the equivalence between
(\ref{mass1}) and (\ref{mass2}) let us
perform the variation of (\ref{mass2})
with respect to $\Phi^A_{ \ B}$
\begin{equation}
\delta^A_{ \ B}- (\Phi^{-1})^A_{ \ C}
\bA^C_{ \ D}
 (\Phi^{-1})^D_{ \ B}=0
\end{equation}
that implies
\begin{equation}\label{PhiABA}
\Phi^A_{ \ B}\Phi^{B}_{ \ C}= \bA^A_{ \
C} \Rightarrow \Phi^A_{ \ B}= \sqrt{
\bA^A_{ \ B}} \ .
\end{equation}
Then inserting (\ref{PhiABA}) into
(\ref{mass2}) we obtain (\ref{mass1})
and hence an equivalence between these
two formulations is established.

 Now we are ready
to proceed to the Hamiltonian formalism
of given theory.
 Explicitly,
we use  $3+1$ notation
\cite{Arnowitt:1962hi} \footnote{For
review, see \cite{Gourgoulhon:2007ue}.}
and write the four dimensional metric
components as
\begin{eqnarray}
\hat{g}_{00}=-N^2+N_i g^{ij}N_j \ ,
\quad \hat{g}_{0i}=N_i \ , \quad
\hat{g}_{ij}=g_{ij} \ ,
\nonumber \\
\hat{g}^{00}=-\frac{1}{N^2} \ , \quad
\hat{g}^{0i}=\frac{N^i}{N^2} \ , \quad
\hat{g}^{ij}=g^{ij}-\frac{N^i N^j}{N^2}
\ .
\nonumber \\
\end{eqnarray}
Note also that $4-$dimensional scalar
curvature has following decomposition
\begin{equation}\label{Rdecom}
{}^{(4)}R=K_{ij}\mG^{ijkl}K_{kl}+{}^{(3)}R
\ ,
\end{equation}
where ${}^{(3)}R$ is three-dimensional
spatial curvature, $K_{ij}$ is
extrinsic curvature defined as
\begin{equation}
K_{ij}=\frac{1}{2N} (\partial_t g_{ij}
-\nabla_i N_j-\nabla_j N_i) \ ,
\end{equation}
where $\nabla_i$ is covariant
derivative built from the metric
components $g_{ij}$. Note also that
$\mG^{ijkl}$ is de Witt metric defined
as
\begin{equation}
\mG^{ijkl}=\frac{1}{2}(g^{ik}g^{jl}+g^{il}g^{jk})-g^{ij}
g^{kl} \ , \quad
\mG_{ijkl}=\frac{1}{2}
(g_{ik}g_{jl}+g_{il}g_{jk})-\frac{1}{2}g_{ij}g_{kl}
\ .
\end{equation}
Finally note that in (\ref{Rdecom}) we
omitted terms proportional to the
covariant derivatives which induce the
boundary terms that vanish for suitable
chosen boundary conditions. Then in
$3+1$ formalism $\bA^A_{ \ B}$ takes
the form
\begin{equation}
\bA^A_{ \ B}=
-\nabla_n\phi^A\nabla_n\phi_B+g^{ij}\partial_i\phi^A
\partial_j\phi_B \ ,
\quad \nabla_n\phi^A= \frac{1}{N}
(\partial_t\phi^A-N^i\partial_i \phi^A)
\ .
\end{equation}
Using this notation we rewrite the
mass term (\ref{mass2}) into the form
\begin{eqnarray}
\mL_{matt}&=&
-M_p^2m^2 \sqrt{g}N
(\Phi^{AB}\eta_{BA}-(\Phi^{-1})_{AB}\nabla_n\phi^B\nabla_n\phi^A+
(\Phi^{-1})_{AB}V^{BA}) \ , \nonumber \\
V^{BA}&=& g^{ij}\partial_i\phi^B\partial_j\phi^A
\ , \quad  \Phi_{AB}=\eta_{AC}\Phi^C_{ \ A} \
. \nonumber \\
\end{eqnarray}
The Hamiltonian analysis of the gravity
part of the action is well known.
Explicitly, the momenta conjugate to
$N,N^i$ are the primary constraints of
the theory
\begin{equation}
\pi_N(\bx)\approx 0 \ , \quad
\pi_i(\bx)\approx 0 \
\end{equation}
while the Hamiltonian takes the form
\begin{eqnarray}\label{HamGR}
\mH^{GR}&=& N\mH^{GR}_T+N^i\mH^{GR}_i \
,
\nonumber \\
\mH_T^{GR}&=&\frac{1}{\sqrt{g}M_p^2}
\pi^{ij}\mG_{ijkl}\pi^{kl}-M_p^2
\sqrt{g} {}^{(3)}R \ , \nonumber \\
\mH_i^{GR}&=& -2
g_{ik}\nabla_j\pi^{jk} \ , \nonumber \\
\end{eqnarray}
where $\pi^{ij}$ are momenta conjugate
to $g_{ij}$ with following non-zero
Poisson brackets
\begin{equation}
\pb{g_{ij}(\bx),\pi^{kl}(\by)}=
\frac{1}{2}\left(\delta_i^k\delta_j^l+\delta_i^l
\delta_j^k\right)\delta(\bx-\by) \ .
\end{equation}
Finally note that $\pi_N,\pi_i$ have
following Poisson brackets with $N,N^i$
\begin{equation}
\pb{N(\bx),\pi_N(\by)}=\delta(\bx-\by)
\ , \quad \pb{N^i(\bx),\pi_j(\by)}=
\delta^i_j\delta(\bx-\by) \ .
\end{equation}
Now we proceed to the Hamiltonian
analysis of the scalar field part of
the action. The momenta conjugate
 $\Phi^{AB}$ are the primary constraints
\begin{equation}
P_{AB}=\frac{\delta
\mL_{matt}}{\delta
\partial_t\Phi^{AB}}\approx 0 \ ,
\end{equation}
while the momenta conjugate to $\phi^A$
are related to the time derivative of
$\phi^A$ by
\begin{equation}
p_A=\frac{\delta \mL_{matt}}{\delta
\partial_t\phi^A}
=2M_p^2m^2\sqrt{g}(\Phi^{-1})_{AB}
\nabla_n\phi^B \ .
\end{equation}
Note that these canonical variables
 have following
non-zero Poisson brackets
\begin{equation}
\pb{\Phi^{AB}(\bx),P_{CD}(\by)}=
\frac{1}{2}(\delta^A_C\delta^B_D+
\delta^A_D\delta^B_C)\delta(\bx-\by) \
 , \quad  \pb{\phi^A(\bx),p_B(\by)}=\delta^A_B\delta(\bx-\by) \ .
\end{equation}
Using these variables we find that the
 Hamiltonian for the scalar
fields takes the form
\begin{eqnarray}
\mH^{sc}&=&
N\mH^{sc}_T+N^i\mH^{sc}_i+\Omega^{AB}P_{BA}
\ , \quad
\mH^{sc}_i=p_A\partial_i\phi^A \ ,
\nonumber
\\
\mH^{sc}_T&=&\frac{1}{4\sqrt{g}M_p^2m^2}
\Phi^{AB}p_B
p_A+M_p^2m^2\sqrt{g}[\Phi^{AB}\eta_{BA}
+(\Phi^{-1})_{AB}V^{BA}] \
 .  \nonumber \\
\end{eqnarray}
Now we analyze the requirement of the
preservation of the primary constraints
during the time evolution of the system. Note that
the Hamiltonian is equal to
\begin{equation}
H=\int
d^3\bx(N\mH_T+N^i\mH_i+\Omega^{AB}P_{BA}+u^N\pi_N+
u^i \pi_i) \ ,
\end{equation}
where $\mH_T=\mH_T^{GR}+\mH_T^{sc} \ ,
 \mH_i=\mH_i^{GR}+ \mH_i^{sc}$ and
 where
 $\Omega^{AB},u^N$ and $u^i$ are
 Lagrange
 multipliers corresponding to the
 primary constraints $P_{AB}\approx 0,\pi^N\approx 0$ and
 $\pi_i\approx 0$.

As usual the requirement of the
preservation of the primary constraints
$\pi_N\approx 0 , \pi_i\approx 0$
implies following secondary constraints
\begin{equation}
\mH_T\approx 0 \ , \quad  \mH_i\approx
0 \ .
\end{equation}
On the other hand the requirement of
the preservation of the primary
constraints $P_{AB}\approx 0$ implies
following secondary ones
\begin{eqnarray}\label{defPsi}
\partial_t P_{AB}&=&\pb{P_{AB},H}= -\frac{N}{4M_p^2 m^2\sqrt{g}} p_A
p_B+\nonumber \\
&+& NM_p^2m^2\sqrt{g}[-\eta_{BA}+
(\Phi^{-1})_{AC}V^{CD}(\Phi^{-1})_{DB}]
\equiv N\Psi_{AB}\approx 0
\nonumber \\
\end{eqnarray}
using
\begin{eqnarray}
\pb{P_{AB}(\bx),(\Phi^{-1})_{CD}(\by)}&=&
-(\Phi^{-1})_{CK}(\by)\pb{P_{AB}(\bx),\Phi^{KL}(\by)}
(\Phi^{-1})_{LD}(\by)=\nonumber
\\
&=&\frac{1}{2}
((\Phi^{-1})_{AC}(\Phi^{-1})_{BD}+
(\Phi^{-1})_{AD}(\Phi^{-1})_{BC})(\bx)
\delta(\bx-\by) \ .
\nonumber \\
\end{eqnarray}
It is very interesting to analyze the constraint $\Psi_{AB}$
further, following \cite{Hassan:2012qv}. Note that we can express
$V^{AB}$ using (\ref{defPsi}) as
\begin{equation}\label{defVCD}
V^{CD}=\Phi^{CA}\left(\Psi_{AB}+\frac{p_A p_B}{4M_p^2 m^2 \sqrt{g}}
+M_p^2 m^2 \sqrt{g}\eta_{AB}\right)\Phi^{BD} \ .
\end{equation}
By definition $V^{CD}=\partial_i \Phi^A g^{ij}\partial_j \phi^D
\equiv (W^T)^C_i g^{ij} W_j^{  \ D}$ where $W_i^A=\partial_i \phi^A$
is $3\times 4$ matrix and $W^T$ is its transverse. As a result all
matrices on the right side have the rank $3$ and consequently
$V^{CD}$ is the matrix of the rank $3$ too. This fact however
implies that $V^{CD}$ is singular matrix with vanishing determinant.
Then calculating the determinant from (\ref{defVCD}) we obtain (note
that $\Phi^{AB}$ is non-singular matrix by definition)
\begin{equation}
\det\left(\Psi_{AB}+\frac{p_A p_B}{4M_p^2 m^2 \sqrt{g}} +M_p^2 m^2
\sqrt{g}\eta_{AB}\right)\approx \det \left(\frac{p_A p_B}{4M_p^2 m^2
\sqrt{g}} +M_p^2 m^2 \sqrt{g}\eta_{AB}\right)=0 \ .
\end{equation}
In other words we obtain that on the constraint surface
$\Psi_{AB}\approx 0$ following expression vanishes
\begin{eqnarray}
& &\det \left(\frac{p_A p_B}{4M_p^2 m^2 \sqrt{g}} +M_p^2 m^2
\sqrt{g}\eta_{AB}\right)=\nonumber \\
&=& M_p^2m^2\sqrt{g}\left(
 \frac{1}{4M_p^4 m^4 g}p_A p^A+1\right)\equiv M_p^2m^2\sqrt{g}
 \mC\approx 0
\nonumber \\
\end{eqnarray}
which we denote as $\mC\approx 0$. Note that we should interpret it
as a constraint that follows from the fact that $\det V^{AB}=0$.
Note also that using this result we find the eigenvector of the
matrix $V^{AB}$ very easily. In fact,  when we multiply (\ref{defVCD}) from the
right by $u_A\equiv (\Phi^{-1})_{AB}\eta^{BC}p_C$ we obtain
\begin{eqnarray}
& &V^{AB}u_B\approx \Phi^{AC} \left(\frac{1}{4M_p^2 m^2\sqrt{g}}p_C p_B
\eta^{BD}p_D+ M_p^2m^2\sqrt{g}p_C\right)=\nonumber \\
&=& M_p^2m^2\sqrt{g}\Phi^{AC}p_C \left(\frac{1}{4M_p^4
m^4g}p_Ap^A+1\right)=
 M_p^2m^2\sqrt{g}\Phi^{AC}p_C\mC\approx 0 \ . \nonumber \\
\end{eqnarray}
For further purposes it is useful to know the Poisson bracket
between $P_{AB}$ and  $\Psi_{CD}$
\begin{eqnarray}
\pb{P_{AB}(\bx),\Psi_{CD}(\by)} &=&
\frac{1}{2}M_p^2m^2
\sqrt{g}[((\Phi^{-1})_{AC}(\Phi^{-1})_{BE}+
(\Phi^{-1})_{AE}(\Phi^{-1})_{BC})V^{EF}
(\Phi^{-1})_{FD}+
\nonumber \\
&+&(\Phi^{-1})_{CE}V^{EF}
((\Phi^{-1})_{AF}(\Phi^{-1})_{BD}+
(\Phi^{-1})_{AD}(\Phi^{-1})_{BF})
]\delta(\bx-\by)\equiv\nonumber \\
&\equiv& \triangle_{AB,CD}(\bx)
\delta(\bx-\by) \ . \nonumber \\
\end{eqnarray}
Note that there are  non-trivial
Poisson brackets $
\pb{\Psi_{AB}(\bx),\Psi_{CD}(\by)}$
and
$\pb{\Psi_{AB}(\bx),\mH_T(\by)}\equiv
\triangle_{AB}(\bx,\by)$. Fortunately
their explicit forms will not be
needed.
Finally note that it is useful to write
the total Hamiltonian in the form
\begin{eqnarray}
H_T&=&\int d^3\bx( N\mH_T+\Omega^{AB}P_{BA}+\Gamma^{AB}
\Psi_{BA}+\Sigma \mC)+\bT_S(N^i) \ , \nonumber
\\
\bT_S(N^i)&=&\int d^3\bx N^i(\mH_i+\partial_i\Phi^{AB}P_{BA}) \ ,
\nonumber \\
\end{eqnarray}where $\Omega^{AB},\Gamma^{AB}$ and $
\Sigma$ are corresponding Lagrange multipliers.
 Using the smeared
form of the diffeomorphism constraint $\bT_S(N^i)$ we find the
Poisson bracket
\begin{equation}
\pb{\bT_S(N^i),\Sigma_{AB}}=
-\partial_i
N^i\Sigma_{AB}-N^i\partial_i\Sigma_{AB}
\ .
\end{equation}
Then it is easy to determine the time
evolution of all constraints. Note also
that the Poisson brackets between
smeared form of the
Hamiltonian and diffeomorphism  constraints
take the standard form
\begin{eqnarray}\label{pbHDC}
\pb{\bT_T(N),\bT_T(M)}&=&
\bT_S((N\partial_jM-M\partial_j
N)g^{ji}) \ ,
\nonumber\\
\pb{\bT_S(N^i),\bT_S(M^j)}&=&
\bT_S(N^j\partial_j M^i-M^j\partial_j
N^i) \ , \nonumber \\
\pb{\bT_S(N^i),\bT_T(N)}&=&
\bT_T(\partial_k N N^k) \ .
\nonumber \\
\end{eqnarray}
In case of the  constraints $P_{AB}$
and $\Psi_{AB}$ we obtain
\begin{eqnarray}\label{evolPPsi}
\partial_t P_{AB}&=&\pb{P_{AB},H_T}\approx
\triangle_{AB,CD}\Gamma^{CD}=0 \ ,
\nonumber\\
\partial_t\Psi_{AB}&\approx&
\int d^3\bx \left(\pb{\Psi_{AB},N\mH_T(\bx)}
+\Gamma^{CD}\pb{\Psi_{AB},\Psi_{CD}(\bx)}+\right. \nonumber \\
&+&\left. \Omega^{CD}\pb{\Psi_{AB},P_{CD}(\bx)}+\Sigma
\pb{\Psi_{AB},\mC(\bx)}\right)=0 \ .
\nonumber \\
\end{eqnarray}
The crucial point is to find the non-trivial
 solution of the equation
\begin{equation}\label{triangleGamma}
\triangle_{AB,CD}\Gamma^{CD}=0 \ .
\end{equation}
Let us consider  following  ansatz
\begin{equation}
\Gamma^{AB}=\Phi^{AC}u_C\Phi^{BD} u_D=\Gamma^{BA} \ ,
\end{equation}
where $u_A=(\Phi^{-1})_{AB}\eta^{BC}p_C$. Inserting this ansatz into
the equation (\ref{triangleGamma}) we obtain
\begin{eqnarray}
\triangle_{AB,CD}\Gamma^{CD}
=M_p^2m^2 \sqrt{g} [u_A (\Phi^{-1})_{BE}V^{EF}u_F+
(\Phi^{-1})_{AE}V^{EF}u_Fu_B]=0 \ .
\nonumber \\
\end{eqnarray}
Let us then consider following linear
combination
\begin{equation}
P^M=\Phi^{AC}u_C\Phi^{BD}u_D P_{AB} \ .
\end{equation}
Then  we can write $P_{AB}$ as
\begin{equation}\label{splitP}
P_{AB}=\tilde{P}_{AB}+\frac{1}{(u_A u^A)^2} P^M (\Phi^{-1})_{AC}u^C
(\Phi^{-1})_{BD}u^D \ ,
\end{equation}
where $\tilde{P}_{AB}$ obey by definition following condition
\begin{equation}
\tilde{P}_{AB}(\Phi^{-1})_{AC}u^C(\Phi^{-1})_{BD}u^D=0 \ .
\end{equation}
Let us now  consider the constraint
\begin{equation}\label{defPsiM}
\Psi_M=\Psi_{AB}\Phi^{AC}u_C\Phi^{BD}u_D= -M_p^2m^2 \sqrt{g}p_Ap^A
\left(\frac{1}{4M_p^4 m^4g}p_Ap^A+1\right)= -M_p^2 m^2 \sqrt{g}\mC \
.
\end{equation}
using the fact that $V^{AB}u_B\approx 0$. Then we see that when we
split $\Psi_{AB}$ as
\begin{equation}\label{splitPsi}
\Psi_{AB}=\tilde{\Psi}_{AB}+ \frac{1}{(u_A u^A)^2}
(\Phi^{-1})_{AC}u^C (\Phi^{-1})_{BD}u^D \Psi_M
\end{equation}
we find that
\begin{equation}
\tilde{\Psi}_{AB}(\Phi^{-1})_{AC}u^C(\Phi^{-1})_{BD}u^D=0 \ .
\end{equation}
In other words $\tilde{\Psi}_{AB}$ contain $9$ independent
constraints. Now we see that it is natural to consider the total
Hamiltonian in the form
\begin{equation}\label{HTnew}
H_T=\int d^3\bx( N\mH_T+\tilde{\Omega}^{AB}\tilde{P}_{BA}+\Omega_M
P_M+\tilde{\Gamma}^{AB} \tilde{\Psi}_{BA}+\Sigma \mC)+\bT_S(N^i) \ ,
\end{equation}
where we used the split of the constraints
(\ref{splitP}),(\ref{splitPsi}) and where the Lagrange multiplier
corresponding to the constraint $\Psi_M$ was absorbed to the
Lagrange multiplier $\Sigma$ using (\ref{defPsiM}).

Let us now check the requirement of the preservation of all
constraints using the Hamiltonian (\ref{HTnew}).
 Note that $\tilde{\Gamma}^{AB}$ and $\tilde{\Omega}^{AB}$ have to
obey the conditions
\begin{equation}\label{tildeGO}
\tilde{\Gamma}^{AB}\neq (\Phi^{-1})_{AC}u^C(\Phi^{-1})_{BD}u^D \ ,
\quad \tilde{\Omega}\neq (\Phi^{-1})_{AC}u^C(\Phi^{-1})_{BD}u^D \ .
\end{equation}
 In case of the constraints $P_M,\tilde{P}_{AB}$ we
obtain
\begin{equation}
\partial_t P^M=
\pb{P^M,H_T}\approx 0 \
\end{equation}
using $\pb{P^M,\Psi_{AB}}\approx 0$. Then we see that
 $P^M$ is the first
class constraint  of the theory. On the other hand in case of
$\tilde{P}_{AB}$ we find
\begin{equation}
\partial_t \tilde{P}_{AB}\approx
\int d^3\bx \tilde{\Gamma}^{CD} \pb{\tilde{P}_{AB},
\tilde{\Psi}_{CD}}\approx \int d^3\bx \tilde{\Gamma}^{CD}
\triangle_{AB,CD}=0 \ .
\end{equation}
Then due to the condition (\ref{tildeGO}) we obtain that the only
solution of given equation is $\tilde{\Gamma}^{CD}=0$.

Now we proceed to the analysis of the time evolution of the
constraint $\tilde{\Psi}_{AB}$
\begin{eqnarray}
\partial_t\tilde{\Psi}_{AB}&\approx&
\int d^3\bx \left(\pb{\tilde{\Psi}_{AB},N\mH_T(\bx)}
+\tilde{\Gamma}^{CD}\pb{\tilde{\Psi}_{AB},\tilde{\Psi}_{CD}(\bx)}+\right. \nonumber \\
&+&\left.
\tilde{\Omega}^{CD}\pb{\tilde{\Psi}_{AB},P_{CD}(\bx)}+\Sigma
\pb{\tilde{\Psi}_{AB},\mC(\bx)}\right)= \
\nonumber \\
&=&\int d^3\bx \left(\pb{\tilde{\Psi}_{AB},N\mH_T(\bx)}-
\tilde{\Omega}^{CD}\triangle_{CD,AB}+\Sigma
\pb{\tilde{\Psi}_{AB},\mC(\bx)}\right)=0 \ . \nonumber \\
\end{eqnarray}
Now since again $\tilde{\Omega}^{CD}$ have to obey (\ref{tildeGO})
we can presume that given equation can be solved for
$\tilde{\Omega}^{CD}$ at lest in principle. Finally we proceed to
the requirement of the preservation of the constraint $\mC$
during the time evolution of the system
\begin{eqnarray}\label{partSigmaM}
\partial_t \mC&\approx &\pb{\mC,H_T}=
\int d^3\bx \left(N(\bx)\pb{\mC,\mH_T(\bx)}+
\pb{\mC,\tilde{\Psi}_{AB}(\bx)}\tilde{\Gamma}^{AB}(\bx)+
\right.\nonumber \\
&+& \left.\pb{\mC,\tilde{P}_{AB}(\bx)} \tilde{\Omega}^{AB}(\bx)+
\pb{\mC,P_M(\bx)}\Omega_M\right)=
 \int d^3\bx (N\pb{\mC,\mH_T})=0
\nonumber \\
\end{eqnarray}
using the fact that $\tilde{\Gamma}^{AB}=0$ and the fact that
$\pb{\mC,\tilde{P}_{AB}}=\pb{\mC,P_M}=0$. To proceed further we have
to calculate the Poisson bracket between $\mC$
 and $\mH_T$
\begin{eqnarray}\label{SigmaMmHT}
&&\pb{\mC(\bx),\mH_T(\by)}
=-\frac{1}{4M_p^4m^4g^{3/2}} g_{ij}\pi^{ji} p_A p^A\delta(\bx-\by)-
\nonumber \\
&-& \frac{p^ A(\bx)}{2M_p^2m^2 g(\bx)}
(\Phi^{-1})_{AB}\partial_{y^i}\phi^B
g^{ij}(\by)\partial_{y^j}\delta(\bx-\by) \ , \nonumber \\
\end{eqnarray}
using
\begin{equation}
\pb{g(\bx),\mH_T^{GR}(\by)}=
\frac{\sqrt{g}}{M_p^2m^2}g_{ij}\pi^{ji}(\bx) \delta(\bx-\by) \ .
\end{equation}
Then when we insert (\ref{SigmaMmHT}) into (\ref{partSigmaM}) we
obtain
\begin{eqnarray}\label{partmC}
\partial_t \mC=
-\frac{1}{4M_p^4m^4g^{3/2}} g_{ij}\pi^{ji} p_A p^A N+\nonumber \\
+\partial_j N \frac{p^A}{2M_p^2m^2 g}
(\Phi^{-1})_{AB}\partial_{i}\phi^B g^{ij}+ N\frac{p^A}{2M_p^2m^2 g}
\partial_i\left[\sqrt{g}(\Phi^{-1})_{AB}\partial_{j}\phi^B g^{ij}\right]
\nonumber \\
\end{eqnarray}
Observe that the expression proportional to $\partial_i N$ vanishes
on the constraint surface. This follows from the fact that when we
multiply $\Psi_{AB}$ with $p^B$ and with $p_A$  we obtain
\begin{eqnarray}
\Psi_{AB}p^B=M_p^2 m^2 p_A\sqrt{g}\mC+M_p^2m^2
p^A(\Phi^{-1})_{AC}\partial_i\phi^C g^{ij} \partial_j \phi^D
(\Phi^{-1})_{DB}p^B\approx 0 \nonumber \\
\end{eqnarray}
that implies
\begin{equation}
p^A(\Phi^{-1})_{AB}\partial_i\phi^B\approx 0 \ .
\end{equation}
As a result (\ref{partmC}) takes the form
\begin{eqnarray}\label{partmCF}
\partial_t \mC=N\left(
-\frac{1}{4M_p^4m^4g^{3/2}} g_{ij}\pi^{ji} p_A p^A +
\frac{p^A}{2M_p^2m^2 \sqrt{g}}
g^{ij}\partial_i\left[(\Phi^{-1})_{AB}\partial_{j}\phi^B
\right]\right)
\equiv N\mC^{II}=0 \ . \nonumber \\
\end{eqnarray}
We have proceed to the important result that deserves careful
treatment. The issue is that the lapse function $N$ is the Lagrange
multiplier. Then if  $\mC^{II}$ would be non-zero over the whole
phase space we could certainly impose the condition
$N=0$.
Equivalently, since the matrix $\pb{\mC(\bx),\mH_T(\by)}$ is
invertible on the whole phase space we have that $\mC$ and $\mH_T$
could be interpreted as the second class constraints. However the
problem is that the matrix $\pb{\mC(\bx),\mH_T(\by)}$ depends on the
phase space variables and hence it can vanish at different points of
the phase space. For that reason it is not quiet right to say that
this is invertible matrix and then say that $\mH_T,\mC$ are the
second class constraints. For that reason it is more natural to
interpret $\mC^{II}$ as the additional constraint. Then we still
have that $\mH_T$ is the first class constraint and hence we still
have the theory with four first class constraints that is the
manifestation of the fact that we have fully diffeomorphism
invariant theory. On top of that we have two constraints
$\mC,\mC^{II}$ with following non-zero Poisson brackets
\begin{eqnarray}\label{mCmCII}
\pb{\mC(\bx),\mC^{II}(\by)}&=&3\frac{(p_Ap^A)^2}{(4M_p^4m^4)^2
g^{5/2}}(\bx)\delta(\bx-\by)-\nonumber
\\
&-&\frac{p^A(\bx)}{4M_p^4m^4g(\bx)}
p^B(\by)\frac{g^{ij}(\by)}{M_p^2m^2\sqrt{g}(\by)}
\partial_{y^i}[(\Phi^{-1})_{BA}\partial_{y^j}\delta(\bx-\by)] \ .
 \nonumber \\
\end{eqnarray}
Now we should check the stability of all constraints including
$\mC^{II}$. Explicitly we consider now the extended Hamiltonian
\begin{equation}\label{HTnew2}
H_T=\int d^3\bx( N\mH_T+\tilde{\Omega}^{AB}\tilde{P}_{BA}+\Omega_M
P_M+\tilde{\Gamma}^{AB} \tilde{\Psi}_{BA}+\Sigma
\mC+\Sigma^{II}\mC^{II})+\bT_S(N^i) \ .
\end{equation}
In case of $P_M$ we find
\begin{eqnarray}
\partial_t P_M=\pb{P_{M},H_T}\approx
\int
d^3\bx\left(\tilde{\Gamma}^{CD}\pb{P_M,\tilde{\Psi}_{CD}(\by)}+
\Sigma^{II}\pb{P_M,\mC^{II}(\by)}\right) \nonumber \\
\end{eqnarray}
using
 \begin{eqnarray}
\pb{P_{M}(\bx),\mC^{II}(\by)}&=&
\pb{P_{M}(\bx),\frac{1}{N(\by)}
\pb{\mC(\by),\bT_T(N)}}=
\nonumber \\
&=&-\frac{1}{N(\by)}
\pb{\mC(\by),\pb{\bT_T(N),P_M(\bx)}}-
\frac{1}{N(\by)}
\pb{\bT_T(N),
\pb{\mC(\by),P_M(\bx)}}\sim
\nonumber \\
&\sim & -\pb{\mC(\bx),\mC(\by)}=0 \ .  \nonumber \\
\end{eqnarray}
In other words $P_M$ is still the first class constraint.
In case of $\tilde{P}_{AB}$ we obtain that the requirement of the
preservation of given constraint takes the form
\begin{eqnarray}\label{timePABn}
\partial_t \tilde{P}_{AB}=\pb{\tilde{P}_{AB},H_T}\approx
\int
d^3\bx\left(\tilde{\Gamma}^{CD}\pb{\tilde{P}_{AB},\tilde{\Psi}_{CD}(\by)}+
\Sigma^{II}\pb{\tilde{P}_{AB},\mC^{II}(\by)}\right)=0 \nonumber \\
\end{eqnarray}
that implies that now $\tilde{\Gamma}^{AB}$ is not equal to zero.
In fact, due to the restriction on the Lagrange multipliers $\tilde{\Gamma}^{CD}$
(\ref{tildeGO}) we now find that (\ref{timePABn}) can be solved for
$\tilde{\Gamma}^{AB}$ at least in principle.
Further, the requirement of the preservation of  the constraint $\mC$ gives
\begin{eqnarray}\label{timemCnew}
\partial_t \mC=\pb{\mC,H_T}
&\approx& \int d^3\bx \left(N(\bx)\pb{\mC,\mH_T(\bx)}+
\tilde{\Gamma}^{AB}\pb{\mC,\tilde{\Psi}_{AB}(\bx)}+
\Sigma^{II}\pb{\mC,\mC^{II}(\bx)}\right)\approx \nonumber \\
&\approx& \int d^3\bx\left(
\tilde{\Gamma}^{AB}\pb{\mC,\tilde{\Psi}_{AB}(\bx)}+
\Sigma^{II}\pb{\mC,\mC^{II}(\bx)}\right)=0 \ .  \nonumber \\
\end{eqnarray}
Now inserting $\tilde{\Gamma}^{AB}$ derived from (\ref{timePABn}) we find that
(\ref{timemCnew}) is the linear equation for $\Sigma$ that has trivial solution
$\Sigma^{II}=0$. Then however (\ref{timePABn}) implies $\tilde{\Gamma}^{AB}=0$. Finally we
analyze the time evolution of the constraint $\mC^{II}$
\begin{eqnarray}
\partial_t \mC^{II}=\pb{\mC^{II},H_T}=
\int d^3\bx \left(N(\bx)\pb{\mC^{II},\mH_T(\bx)}+
\Sigma (\bx)\pb{\mC^{II},\mC(\bx)}\right)=0
\nonumber \\
\end{eqnarray}
that using (\ref{mCmCII}) we see that given equation reduces to the
linear differential equation for $\Sigma$ that could be solved at least in principle.
Note also that $\mH_T$ is preserved due to the fact that $\Sigma^{II}=0$.

As a result we obtain following picture. We have theory with eight first class constraints
$p_N\approx 0 \ , p^i\approx 0, \tmH_T\approx 0 \ , \tmH_i\approx 0$ and $10$
gravity degrees of freedom $N,N_i,g_{ij}$ together with corresponding conjugate momenta. By gauge
fixing of all constraints we find four phase space degrees of freedom corresponding
to the massless gravity. We have further one first class constraint $P_M\approx 0$.
By gauge fixing of given constraint together with
  $18$ second class constraints $\tilde{P}_{AB},\tilde{\Psi}_{AB}$
and also with two second class constraints $\mC\approx 0 \ , \mC^{II}\approx 0$
we can eliminate all auxiliary degrees of freedom together with one scalar
degree of freedom. As a result we have $6$ phase space degrees of freedom
that together with $4$ degrees of freedom of the gravity sector correspond to the correct
number of the degrees of freedom of massive gravity. In other words we proved the
non-perturbative consistency of the particular model of the non-linear massive
gravity written with the St\"{u}ckelberg fields.
Our result is in the full agreement with the analysis presented in
\cite{Hassan:2012qv}.



 \noindent {\bf
Acknowledgements:}
 This work   was
supported by the Grant agency of the
Czech republic under the grant
P201/12/G028. \vskip 5mm

\end{document}